\documentclass[twocolumn,preprintnumbers,superscriptaddress]{revtex4}
\usepackage{graphicx}
\pdfoutput=1

\usepackage{hyperref}
\usepackage{amsmath,amssymb,bm,bbm,amsfonts}
\usepackage{graphicx}
\usepackage{yfonts}
\usepackage{enumerate}
\usepackage{color}

\newcommand{\bea}{\begin{eqnarray}}
\newcommand{\eea}{\end{eqnarray}}

\newcommand{\SU}{\text{SU}}
\newcommand{\SO}{\text{SO}}

\newcommand{\U}{\text{U}}
\newcommand{\id}{\mathbbm{1}}

\begin{document}

\preprint{OU-HET-783, RIKEN-MP-69, YITP-13-22}

\title{Towards Holographic Spintronics}

\author{Koji Hashimoto}
\email{koji@phys.sci.osaka-u.ac.jp}
\affiliation{Department of Physics, Osaka University, Toyonaka, Osaka 560-0043, JAPAN}
\affiliation{Mathematical Physics Laboratory, 
RIKEN Nishina Center, Saitama 351-0198, JAPAN}

\author{Norihiro Iizuka}
\email{iizuka@yukawa.kyoto-u.ac.jp}
\affiliation{Department of Physics, Osaka University, Toyonaka, Osaka 560-0043, JAPAN}
\affiliation{{Yukawa Institute for Theoretical Physics, 
Kyoto University, Kyoto 606-8502, JAPAN}}

\author{Taro Kimura}
\email{taro.kimura@riken.jp}
\affiliation{Institute de Physique Th\'eorique, CEA Saclay, 91191 Gif-sur-Yvette, FRANCE}
\affiliation{Mathematical Physics Laboratory, 
RIKEN Nishina Center, Saitama 351-0198, JAPAN}

\begin{abstract}
We study transport phenomena of total angular momentum in holography, 
as a first step toward holographic understanding of spin transport phenomena.
Spin current, which has both the local Lorentz index for spins and the space-time vector index for current, 
couples naturally to the bulk spin connection.
Therefore the bulk spin connection becomes the source for the boundary spin current.
This allows us to evaluate the spin current holographically, with a relation to 
the stress tensor and metric fluctuations in the bulk. 
We examine the spin transport coefficients and the thermal spin Hall conductivity 
in a simple holographic setup.
%
\end{abstract}

\maketitle


\,\,\,
\centerline{\bf Introduction}

{\em Spintronics} is a technology where we manipulate the intrinsic
electron spin degrees of freedom instead of the electric
charge~\cite{Maekawa:2006OUP,Zutic:2011NM}. 
In ferromagnetic/anti-ferromagnetic materials, spin-charge separation 
can occur and in such a situation, it is useful to consider spin
as an independent degree of freedom which carries information.
Because electric charge transport is not involved there,
spin devices can reduce power consumption compared to usual
electric ones, and exceed the velocity limit of the electron charge.
This spintronics is actually used widely, for example, for 
read-heads of hard-drives, and is based on a recent development of 
experimental technologies manipulating imbalance between up-spins and down-spins.
For these reasons, spin transport phenomena have been attracting special
interest recently.

Recent research on the spin transport basically relies on
one-body quantum mechanical analyses, especially in the presence of
a spin-orbit interaction.
However, in strongly correlated systems, we have to go beyond the
one-body physics by treating the interaction effect seriously.
In this paper, we propose a method to study the spin transport
phenomena for strongly correlated systems
by using the holography, {\it i.e.,} gauge/gravity
correspondence~\cite{Maldacena:1997re,Gubser:1998bc,Witten:1998qj}.
The method of holography is one of the most useful tools to study strongly
correlated quantum field theories.
While there are some attempts to include effects of spins in 
holography, {\it e.g.}, Refs.~\cite{Benini:2010pr,Harrison:2011fs,Bigazzi:2011ak,Herzog:2012kx,Benincasa:2012wu,PhysRevB.86.125145,Luo:2012am,Hashimoto:2012pb,Ishii:2012hw},
study of spin transport itself has not yet been performed in the literature.
To discuss the spin degrees of freedom,
we first show a definition of spin 
current from a relativistic field theoretical viewpoint as a conserved N\"other's current.
Then with this definition, we show how to deal with the spin
transport coefficients from the holographic viewpoint. 
The key point is that the {\it spin connection} is naturally regarded as a source for the
spin current. 
We demonstrate a holographic
treatment of the spin transport, 
on a ``boosted'' Schwarzschild black brane background in AdS, 
and we calculate a spin transport coefficient and a thermal spin Hall conductivity.


\,\,\,\,

\centerline{\bf Spin current}  

\begin{figure}[t]
 \begin{center}
  \includegraphics[width=25em]{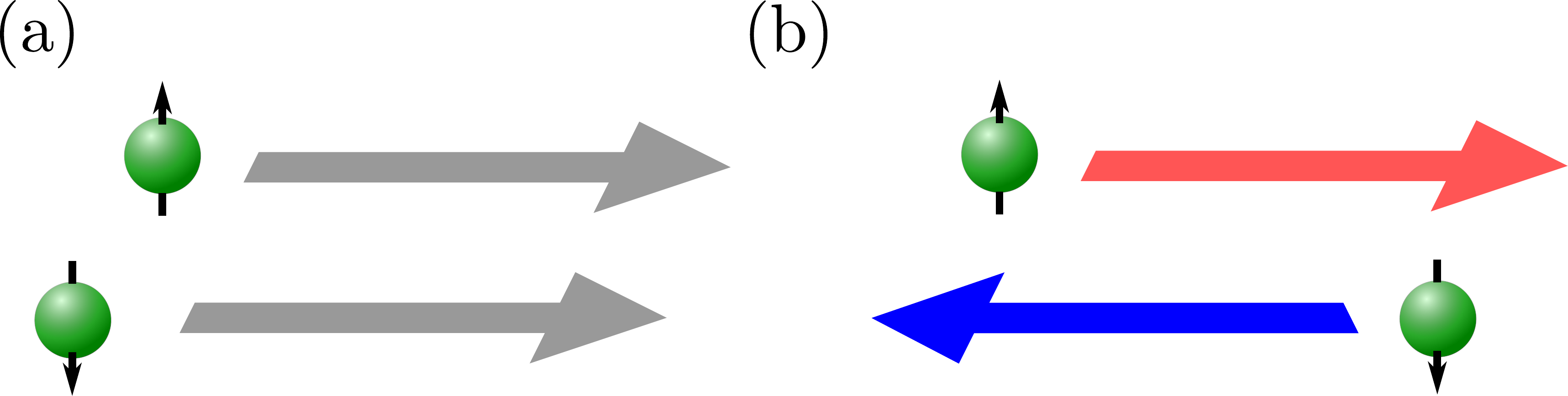}
 \end{center}
 \caption{(a) The charge current is just the total contribution of up-
 and down-spin currents $\vec{J}=\vec{J}_\uparrow+\vec{J}_\downarrow$.
 (b) The spin current is given by difference between them, $\vec{J}_z =
 \frac{1}{2} ( {\vec{J}_\uparrow - \vec{J}_\downarrow} )$. This picture is available if
 and only if $z$-direction spin is conserved.}
 \label{spin_current_fig}
\end{figure}

The spin current is, as the name suggests, a flow of the intrinsic spin
degrees of freedom, instead of the electric charge.
If $z$-spin is conserved, namely a good quantum number, we can
apply a naive definition of the spin current,
\begin{equation}
 \vec{J}_z = \frac{1}{2} 
  \left(
   \vec{J}_\uparrow - \vec{J}_\downarrow
  \right) .
  \label{spin_curr_naive}
\end{equation}
This means that the spin current is given by the difference between
flows of up- and down-spins, $\vec{J}_\uparrow$ and $\vec{J}_\downarrow$,
while the electric current is the total contribution of them,
$\vec{J}=\vec{J}_\uparrow+\vec{J}_\downarrow$, as shown in
Fig.~\ref{spin_current_fig}.
This definition (\ref{spin_curr_naive}) corresponds to the Schwinger
representation of the spin operator, $\vec{s} = \frac{1}{2} \psi^\dag
\vec{\sigma} \psi$.

The expression (\ref{spin_curr_naive}) is 
available if and only if the spin is conserved or,
at least, approximately conserved 
\footnote{For the spin to be approximately conserved, its coherence time
must be sufficiently larger than its characteristic time scale.}. 
However generically the electron spin is not conserved by itself, due to
the spin-orbit interaction. 
Therefore the naive definition of the spin current
(\ref{spin_curr_naive}) has to be modified in the presence of such an
effect.

First we consider how to define the spin current from the field
theoretical point of view.
Let us recall the treatment of conserved currents in
the context of quantum field theories.
A conserved current is defined as a variation of an
action with respect to the corresponding source. 
For example, the electric current $J^\mu$ is derived by differentiating
an action with respect to a $\U(1)$ gauge field,
\begin{equation}
\label{UoneJA}
 J^\mu = \frac{\delta S}{\delta A_\mu} .
\end{equation}
Conservation of $J^\mu$ is guaranteed by N\"other's theorem,
associated with a $\U(1)$ gauge symmetry,
\begin{equation}
 \partial_\mu J^\mu = 0 .
\end{equation}
In the weak coupling limit of a $\U(1)$ gauge theory, the $\U(1)$
local symmetry reduces to a global one.
The $A_\mu$ becomes a non-dynamical background gauge potential, which is a source, 
and the
$J^\mu$ becomes a global current. 
In this limit, the global $\U(1)$ current $J^\mu$ couples to the source
$A_\mu$ in the Lagrangian as ${\cal L}_{\rm source} = A_\mu J^\mu$. 
Therefore, the $\U(1)$ current $J^\mu$ is obtained by 
differentiating the action with respect to its source $A_\mu$. 

Similarly a stress tensor is given by a variation of an action with respect to a
metric,
\begin{equation}
 T^{\mu\nu} = \frac{1}{\sqrt{-g}} \frac{\delta S}{\delta g_{\mu\nu}} .
\end{equation}
The conservation of energy and momentum
\bea
\partial_\mu T^{\mu\nu} = 0 \,
\eea
comes from the
translation invariance in temporal and spatial directions, respectively. 
In the weak gravity limit (where gravity is decoupled),  
non-dynamical background metric $g_{\mu\nu}$ 
becomes a source for the stress-tensor, and they couple as ${\cal L}_{\rm source} 
= g_{\mu\nu} T^{\mu\nu}$ 
in the Lagrangian.

In this way, 
in order to obtain a conserved quantity, we have to
introduce a corresponding field (or source) which couples to the 
conserved quantity. 
For the case of the spin current $J^\mu_{\,\,\, \hat{a}\hat{b}}$, 
our claim is that the spin connection $ \omega_\mu^{\,\,\, \hat{a}\hat{b}}$ is the corresponding field (source). 
This implies that they couple as ${\cal L}_{\rm source} =
  \omega_\mu^{\,\,\, \hat{a}\hat{b}} J^\mu_{\,\,\, \hat{a}\hat{b}}$ 
in the Lagrangian.  
By differentiating an action with respect to the spin connection, 
we can obtain the spin current.

To see why it is so, let us recall the nature of spin. 
The spin operator $s_{\hat a} =  \sigma_{\hat a}/2 $ has an index $\hat a $ for
the orientation of the spin. 
Here the hatted index $\hat a$ takes only a spatial coordinate as
$\hat a = \hat x, \hat y, \hat z$ and $\sigma$ is the Pauli matrix. 
Spin is conserved only in the sense that 
the total angular momentum is conserved. 
The 
total angular momentum is associated with the global rotational symmetry of the system. 
If we uplift this global rotational symmetry to a local one, then
these become a subgroup of the local Lorentz symmetry. 
Therefore, it is natural to associate the conserved spin
$\sigma_{\hat a}$ to a local Lorentz generator $\Sigma_{\hat{a}\hat{b}}=
\frac{i}{4}[\gamma_{\hat{a}},\gamma_{\hat{b}}]$ as  
$\sigma^{\hat a} = \epsilon^{\hat a \hat b \hat c} \Sigma_{\hat{b}\hat{c}}$, where $\epsilon^{\hat a \hat b \hat c} $ 
is anti-symmetric tensor taking $\pm 1$ defined on the spatial part of the local Lorentz indices, {\it i.e.,}
$\hat a, \hat b, \hat c$ of $\epsilon^{\hat a \hat b \hat c} $  takes only $\hat x, \hat y, \hat z$. 
Furthermore, since the spin connection $ \omega_\mu^{\,\,\, \hat{a}\hat{b}}$ is a 
gauge field 
associated with the local Lorentz symmetry, it is natural to associate it  to the conserved spin current $J^\mu_{\,\,\, \hat{a}\hat{b}}$, as equation (\ref{UoneJA}).

Therefore, we reach a conclusion that a spin current is given by a variation of
an action with respect to a spin connection as 
\begin{equation}
 J^\mu_{\,\,\, \hat{a}\hat{b}} = \frac{\delta S}{\delta
  \omega_\mu^{\,\,\, \hat{a}\hat{b}}} \,.
  \label{spin_curr1}
\end{equation}
From now on, the hatted indices $\hat{a}, \hat{b}, \cdots$ represent the local Lorentz indices, so
they stand for $\hat t, 
\hat x, \hat y, \hat z$.
Greek indices $\mu, \nu, \cdots$
stand for curved spacetime vector indices. 
The spin connection is written in terms of a vielbein $e_\mu^{\,\,\, \hat a}$ as 
\bea
\omega_\mu^{\,\,\, \hat a \hat b}  &=& 
e^{\,\,\, \hat{a}}_{\nu} \nabla_\mu e^{\nu \hat{b}} 
\, = \, e^{\,\,\, \hat{a}}_{\nu} \partial_\mu e^{\nu \hat{b}}  
 + e^{\,\,\, \hat{a}}_{\lambda} \Gamma^{\lambda}_{\mu\nu} e^{\nu \hat{b}} \nonumber \\
 &=&  -\,e^{\,\,\, \hat{b}}_{\nu} \nabla_\mu e^{\nu \hat{a}}   \, = \, - \, \omega_\mu^{\,\,\, \hat b \hat a} \,,
  \label{spin_con1}
\eea
where $\Gamma^{\lambda}_{\mu\nu}$ stands for the Christoffel symbol,
and the vielbein $e_\mu^{\,\,\, \hat{a}}$ satisfies $g_{\mu\nu} =
\eta_{\hat{a}\hat{b}}\, e_\mu^{\,\,\, \hat{a}} e_\nu^{\,\,\, \hat{b}}$, with the
local Lorentz metric $\eta_{\hat{a}\hat{b}}=\mathrm{diag}(-1,1,1,1)$.

Usually, we call the following current  
as a spin current
\begin{equation}
 J^{\mu\hat{a}} = \epsilon^{\hat{0}\hat{a}\hat{b}\hat{c}} 
  J^\mu_{\,\,\, \hat{b}\hat{c}},
  \label{spin_curr2}
\end{equation}
rather than the former one $ J^\mu_{\,\,\, \hat{a}\hat{b}}$. Here we use the convention
$\epsilon^{\hat{0}\hat{1}\hat{2}\hat{3}} = 1$. 
One can easily see that the definition (\ref{spin_curr2}) is consistent with, for example,
the standard free fermion spin current. To see this, 
let us consider the generic form of a fermionic Lagrangian on a curved space, which is given by
\begin{equation}
 \mathcal{L}_{\rm F} = 
 \bar\psi 
  \left[
   i e^\mu_{\,\,\, \hat{a}} \gamma^{\hat{a}}
   \left(
    \partial_\mu - i A_\mu - \frac{i}{2}
    \omega_\mu^{\,\,\, \hat{a}\hat{b}} \Sigma_{\hat{a}\hat{b}}
   \right)
   - m
  \right]
  \psi \,.
\end{equation}
From this, we have the spin current by differentiating it with the spin
connection,
\begin{equation}
 J^\mu_{\,\,\, \hat{a}\hat{b}} = \frac{1}{2} \bar\psi 
  \gamma^\mu \Sigma_{\hat{a}\hat{b}} \psi 
  \ \longrightarrow \
 J^\mu_{\,\,\, \hat{a}} 
  = \frac{1}{2} \bar\psi \gamma^\mu (\sigma_{\hat{a}}\otimes\id) \psi .
\end{equation}
This is regarded as a current carrying $\hat{a}$-direction spin.
We can see that the zero-th component correctly gives the spin density
\begin{equation}
 J^0_{\,\,\, \hat{a}} 
  = \psi^\dag (s_{\hat{a}} \otimes \id) \psi .
\end{equation}
In this way, we have seen that the definition (\ref{spin_curr2}) is consistent with the
conventional one for the spin current. 
However it is more convenient to consider $J^\mu_{\,\,\, \hat{a}\hat{b}}$ as a spin current,  
since $ J^\mu_{\,\,\, \hat{a}} $ defined in equation (\ref{spin_curr2}) is not local Lorentz invariant tensor. 
This is because 
$\epsilon^{\hat{0}\hat{a}\hat{b}\hat{c}} $ tensor takes explicit index component $\hat 0$. 
%

The conservation of the spin current  $J^\mu_{\,\,\, \hat{a}\hat{b}}$
\bea
\partial_\mu J^\mu_{\,\,\, \hat{a}\hat{b}} = 0
\eea
is associated with the local Lorentz invariance, and the spin current  $J^\mu_{\,\,\, \hat{a}\hat{b}}$ couples to the source term $\omega_\mu^{\,\,\, \hat{a}\hat{b}}$ in the Lagrangian 
as ${\cal L}_{\rm source} =  \omega_\mu^{\,\,\, \hat{a}\hat{b}} J^\mu_{\,\,\, \hat{a}\hat{b}}$. 

Precisely speaking, what we define above is  ``total angular momentum'' current, rather than 
``spin'' current.  
Note that only the total contribution of the angular momentum current, coming from
both the orbital and the spin angular momentum, is conserved.
A difficulty in dealing with spin transport phenomena is in the
definition of the spin current, 
because the intrinsic spin is not conserved solely but rather conserved as a whole angular momentum. 
Therefore the spin current, by itself, cannot be introduced as a conserved N\"other
current at least in the relativistic limit. 
Thus, in this sense, the spin current defined above is slightly different
from the conventional definition of the spin current often used in the
non-relativistic condensed-matter system, which includes the
contribution of only the intrinsic electron spin.

We will also point out that it is possible that the orbital 
contribution gives only a sub-leading contribution, 
in the non-relativistic limit. 
This is because the orbital angular momentum
includes the spatial momentum as $\vec{L}=\vec{x}\times\vec{p}$. 
Thus, by taking an appropriate limit, the spin current, defined as a
conserved one, may provide a good description of the spin transport. 
We will discuss how we take the non-relativistic limit a bit more in
detail in the discussion later. 

There is a number of attempts to define the spin current in the literature.
The original idea of using the spin connection as a source to obtain a spin current is
found in Refs.~\cite{Wen:1992ej,RevModPhys.65.733}, especially in $2+1$
dimensions. 
In Ref.~\cite{Wen:1992ej} the authors treated the space and time
separately and broke the Lorentz invariance explicitly. 
Another attempt to define a spin current is performed by introducing an
$\SU(2)$-valued gauge field, coupled to a spin degrees of freedom, in
addition to a $\U(1)$ electromagnetic
field~\cite{Mineev:JLTP1989,PhysRevLett.62.482,Frohlich:1992CMP}.
This $\SU(2)$ symmetry can be seen as a remnant of the local Lorentz
symmetry, which is decomposed as $\SO(1,3)\cong\SU(2)\times\SU(2)$
in $3+1$ dimensions.
However, since these $\SU(2)$ are not decoupled except for the massless
case, it is difficult to define the spin current as a
conserved current only with the $\SU(2)$ gauge field.
Actually, this $\SU(2)$ symmetry is broken in the presence of the
spin-orbit interaction.

\,\,\,
\centerline{\bf Holography}
Given the spin current definition in terms of spin connection, 
in order to study the spin current by the gauge/gravity duality scheme, we will 
evaluate the fluctuation mode of the spin connection. 
Note that holography induces one extra coordinate, {\it i.e.}, a radial direction. 
So in the gravity side, the local Lorentz index runs as $\hat a = \hat t, \hat x, \hat
y, \hat z$ and $\hat r$. 
Similarly the vector index runs $\mu =t, x, y, z, r$.

Before studying a component of the spin connection corresponding to
a spin current in a spatial direction, we analyze a temporal component of 
a spin current $J_{t}^{\,\,\, \hat{x}\hat{y}}$, as an example. 
This term couples to $\omega_{t}^{\,\,\, \hat{x}\hat{y}}$. 
When the background metric is diagonal, the static contribution is
calculated as
\begin{equation}
 \delta \omega_{t}^{\,\,\, \hat{x}\hat{y}} = 
\frac{1}{2} e^{x\hat{x}} e^{y\hat{y}}
  \Big(
  \partial_y \delta g_{tx} - \partial_x \delta g_{ty}
  \Big).
\end{equation}
Here we apply a gauge choice $e_r^{\,\,\, \hat{a} \neq \hat r}=g_{r \mu \neq r}=0$.  
From the indices, it is clear that this represents a rotation of a metric fluctuation in 
the $xy$-plane.
In terms of the gauge/gravity duality, the non-normalizable mode of this
component is regarded as a {\em chemical potential} for $\hat
z$-component of the total angular momentum, {\it i.e.,} $\omega_{t ~{\rm
(NN)}}^{\hat x \hat y}  = \frac{1}{2} \mu^{\hat z}$, where the index
${\rm (NN)}$ represents the non-normalizable mode \footnote{The factor
${1}/{2}$ is for a convenience due to the definition, equation
(\ref{spin_curr2}).}.
This chemical potential is naively interpreted as the difference between
those for up- and down-spins, $\mu^{\hat{z}} = \frac{1}{2} ({\mu_\uparrow -
\mu_\downarrow})$. The $\hat z$-component spin density
$J_t^{\,\,\, \hat z}$ corresponds to the normalizable mode of
$\omega_{t ~{\rm (N)}}^{\hat x \hat y}$ in the holographic viewpoint, 
where the index ${\rm (N)}$ represents the normalizable mode.

Similarly, let us study a fluctuation of the spin connection along the $x$-spatial direction, 
$\omega_{x}^{\,\,\, \hat{x}\hat{y}}$. 
This corresponds to a spin current
$J_{x}^{\,\,\, \hat{x}\hat{y}} = \frac{1}{2} J_x^{\,\,\, \hat{z}}$, {\it i.e.,} $\hat z$-oriented ``spin'' 
flows along $x$ direction. 
Here we can see that we need to turn on some of the off-diagonal elements of the
background metric, in particular $g_{tx}$ and $g_{ty}$, which 
correspond to non-vanishing off-diagonal contributions of vielbeins,
$e_t^{\,\,\, \hat{x}}$ and $e_t^{\,\,\, \hat{y}}$.
To see this, assuming that the fluctuation depends only on $r$ and $t$ directions, 
we obtain
\begin{eqnarray}
  \delta \omega_{x}^{\,\,\, \hat{x}\hat{y}} 
=  - \frac{1}{2} e^{t\hat{x}} e^{y\hat{y}} \partial_t \delta g_{xy}
  + \frac{1}{2} e^{x\hat{x}} e^{t\hat{y}} \partial_t \delta g_{xx} .
  \label{spin_conn_fluc}
\end{eqnarray}
From this expression one can see that the off-diagonal
components of the metric, $e_t^{\,\,\, \hat{x}}$ and $e_t^{\,\,\, \hat{y}}$, or
equivalently $g_{tx}$ and $g_{ty}$, are required in order to give the spin
current $J_x^{\,\,\, \hat{z}}$.
A physical meaning of this condition is discussed later.

\,\,\,

\centerline{\bf Example: ``boosted'' black brane}
So far we have considered a boundary theory in 
$3 + 1$ ($\hat x, \hat y, \hat z$ and $\hat t$)  dimensions. 
However, even if the boundary theory is 2 + 1-dimensional,  none of our
argument so far is modified since 2+1-dimensional theories still admit a
spin along the ``$z$''-direction;
Here ``$z$''-direction is simply the $(\hat a, \hat b) = (\hat x , \hat y)$ component, 
$J_\mu^{\,\,\, \hat x \hat y}$.
We will conduct a calculation of the spin current in a 
holographic setting, but for simplicity of the  
calculation in the bulk, we consider a bulk theory in 3+1 dimensions, which corresponds to a boundary theory in 2+1 dimensions. 

We demonstrate a calculation of the transport coefficients for spin
with the simplest holographic setup,
{\it i.e.,} pure gravity in 3+1 dimensions,
\begin{equation}
 S = S_{\rm bulk} + S_{\rm boundary} 
  \, ,
  \label{EH_action}
\end{equation}
\begin{align}
 S_{\rm bulk} & = \int d^4x \, \sqrt{-g} \; (R[g] - 2 \Lambda ) \, , 
 \label{bulk_theory}
 \\
 S_{\rm boundary} & = 2  \int d^3 x \, \sqrt{-\gamma} \, \Theta \, ,
 \label{boundary_theory}
\end{align}
where the cosmological constant is $\Lambda=-3$, and $\gamma_{\mu\nu}$ is the
boundary metric, defined by the metric components along the boundary
dimensions. 
$\Theta$ is a scalar defined with the extrinsic curvature $\Theta^{\mu\nu} 
= -\frac{1}{2} \left( \nabla^\mu n^\nu + \nabla^\nu n^\mu \right)$, as $\Theta =
\gamma_{\mu\nu} \Theta^{\mu\nu}$. 
$n^\mu$ is outward unit vector pointing along the radial direction. 
This boundary action is 
to provide a well-defined Dirichlet variational principle. 
In addition, we have to also take into account another counter term,
called the cosmological counter term, which
depends on the intrinsic curvature of the
boundary~\cite{Balasubramanian:1999re}.
Although this counter term is important for the regulation of the
boundary stress tensor, it is known that the correct boundary stress
tensor, involving the contribution from the cosmological counter term,
can be read-off simply from the normalizable modes of the
metric~\cite{deHaro:2000xn}.
As explained later, we will study the spin current in terms of
the stress tensor based on the relation between the spin connection
and the metric, and furthermore we will read-off the boundary stress
tensor from the normalizable modes. Therefore, we just apply the
argument for the stress tensor, instead of taking the variation with
the spin connection without worrying about the cosmological counter
term.

We study metric fluctuations around a {\it ``boosted''}  Schwarzschild
black brane solution in AdS$_4$,
\begin{eqnarray}
 ds^2  & = & - U(r) dt^2 + \frac{1}{U(r)} dr^2+ r^2  dy^2  \nonumber \\
 &&  + (r^2 - a^2 U(r)) dx^2 -  2 a U(r) dt dx
  \label{boosted_AdS_metric}
\end{eqnarray}
with $U(r) = (r^3-r_0^3)/r$. $r=r_0$ is the horizon while $r=\infty$ is the boundary. 
$r_0$ is related to the temperature $T$ as $T = {3 r_0}/{4 \pi}$ 
\footnote{Our ``boost'' is simply a coordinate transformation. Since it is 
different from the Lorentz boost, it does not involve the $\gamma$ factor for a Lorentz transformation, therefore the temperature does not change by this ``boost''.}. 
This metric was obtained 
by a coordinate transformation $ t \to t + a x$ 
on the AdS-Schwarzschild solution, and it suffices our purpose since it includes
the off-diagonal metric element $g_{tx}$.
We can check that this satisfies the Einstein equation
$R_{\mu\nu}-\frac{1}{2}g_{\mu\nu} R + \Lambda g_{\mu\nu} = 0$, 
and is not singular for $|a|<1$, and we can consider $a>0$ without loss of generality.

Let us perform a fluctuation analysis around the background solution.
Fluctuations we consider are
$\delta g_{ty}$ and $\delta g_{xy}$, and we assume the following form for
AC fluctuations,
\begin{eqnarray}
\label{ACflucepsilon}
&& \delta g_{ty} = \delta g_{yt}  
 = \epsilon \, e^{-i\omega t}r^2 f(r) \,, \quad  \\
&& \delta g_{xy} = \delta g_{yx}
 = \epsilon \, e^{-i\omega t}r^2 h(r) \,.
\label{ACflucepsilontwo}
\end{eqnarray}
Then, nontrivial components of the Einstein equation to linear order 
in these fluctuations, $\mathcal{O}(\epsilon)$, are found to be 
just the $ty$-component, the $ry$-component and the $xy$-component. The other components of the Einstein equation turn out to be trivially satisfied. Among the three equations, the $ry$-component provides a constraint
\begin{eqnarray}
 f'(r) = \left(a + \frac{r^3}{a(r_0^3-r^3)}\right)^{-1}  h'(r)  \,.  
\label{hfrel}
\label{Einsry}
\end{eqnarray}
where $'$ is for the $r$-derivative. 
With this relation, the $ty$-component reduces to a simple equation solely for $h(r)$,
\begin{eqnarray}
h(r) + \frac{r^3-r_0^3}{ \omega^2 r^3}
\frac{d}{dr}
\left[
\frac{(r^3-r_0^3)r^4 }{(1-a^2)r^3+a^2 r_0^3}\frac{d}{dr} h(r)
\right]=0.
\label{Einsty}
\end{eqnarray}
Furthermore, the remaining $xy$-component of the Einstein equations also reduces to the same equation 
(\ref{Einsty}). So, we just need to solve the equation (\ref{Einsty}) for $h(r)$, and relate it to $f(r)$ via
the constraint equation (\ref{Einsry}). This equation (\ref{Einsty}), in the limit $a = 0$, coincides with 
the equation for the shear viscosity calculation \cite{Kovtun:2004de}
\footnote{
Note that since these two equations solve all the Einstein equations, 
these two modes, $\delta g_{ty}$ and $\delta g_{xy}$, 
decouple from the other components of the fluctuation. Therefore, this is a consistent truncation
of the whole Einstein equations.
}.

Eq.~(\ref{Einsty}) can be written by a new coordinate $x\equiv r_0/r$ as
\begin{eqnarray}
\frac{\omega^2}{r_0^2} h(x) = x^2 (x^3-1) \frac{d}{dx}
\left[
\frac{1-x^3}{x^2(1-a^2+a^2 x^3)}\frac{dh(x)}{dx}
\right]  . \,\,
\label{eqx}
\end{eqnarray}
The new coordinate $x$ ranging $0\leq x \leq 1$ can make the boundary analysis easier.

Near the horizon $x=1$, we can solve (\ref{eqx}) as
\begin{eqnarray}
h \propto \exp\left(- \frac{i}{3} \frac{\omega}{r_0} \log (1-x) \right) \,,
\label{horizonx}
\end{eqnarray}
which amounts to the in-going boundary condition at the horizon.
Note that the equation of motion (\ref{eqx}) and the in-going boundary condition (\ref{horizonx}) 
depend on $r_0$ only through the combination $\omega/r_0$. Since $T \propto r_0$, the temperature dependence is the same as the $1/\omega$ dependence. 
This is because the background is a finite temperature system of an AdS
space, namely a scale invariant system, and therefore, any non-trivial
dependence comes from only the dimensionless ratio, $\omega/r_0$ 
\footnote{The large $r/r_0 \to \infty$ is equivalent to the $r_0 \to 0$ with $r$ fixed, where the ratio $\omega/r_0 \to \infty$ by fixing $\omega$. This implies that the bulk large (small) $r$ region corresponds to the large (small) $\omega$ in the boundary theory as in usual UV/IR correspondence \cite{Peet:1998wn}.}.    

Near the boundary $x=0$, we have two independent solutions of (\ref{eqx}),
\begin{eqnarray}
h &=& h_0 \left(1 -\frac12 \gamma x^2 -\frac18 \gamma^2 x^4 + \cdots \right), 
 \\
h &=& h_3 \, \, \Bigl( x^3 + \cdots \Bigr),
\label{h3def}
\end{eqnarray}
with $\gamma(\omega,T) \equiv (1-a^2) \omega^2/r_0^2$.
Here $h_0$ and $h_3$ are integration constants. 
We can find that $h_0$ is the non-normalizable mode while $h_3$ is 
the normalizable mode. 
Consider the bulk action, equation (\ref{bulk_theory}), and expand that around $r \to \infty$ in the background equation (\ref{boosted_AdS_metric}), 
with the fluctuation $h(r)$ and $f(r)$. After using the constraint (\ref{Einsry}),
we find, to the quadratic order in $h(r)$, the leading $r$ behavior of
the Einstein action is
\begin{eqnarray}
& &\sqrt{-g}\left[ R[g] - 2 \Lambda \right] |_{r \to \infty} \nonumber \\
& &= (\mbox{background}) -  \frac{ \epsilon^2 \, e^{-2i\omega t} r^4}{2 (1-a^2)}
 h'(r)^2 \,,  
\label{quadrar}
\end{eqnarray}
neglecting the boundary terms. 
From this expression, we confirm that  $h\sim$ const. is 
the 
non-normalizable mode \footnote{Note that terms like $r^4 h'(r)^2$ are equivalent to terms 
like $r^2 h(r)^2$ through the integration by parts.} while $h\sim r^{-3}$ 
is the normalizable mode.

We can also specify the boundary condition for the other fluctuation $f(r)$. 
From (\ref{hfrel}), we obtain
\begin{eqnarray}
f(x) = \int_{1}^x  \frac{a(s^3-1)}{a^2 s^3 - a^2 + 1} \frac{dh(s)}{ds} ds + c,
\label{constantapp}
\end{eqnarray}
where $c$ is an integration constant. Near the horizon $x=1$, 
$h(x)$ approximated as
(\ref{horizonx}) can give an in-going wave for $f(x)$ only if $c=0$. 
So we need to put $c=0$, 
and $f(x)$ is uniquely determined once $h(x)$ is given.
The magnitude $f_0$ of the 
non-normalizable mode
of $f(x)$ can be read by
(\ref{constantapp}) with $c=0$, while the magnitude $h_3$ of the 
normalizable mode of $f(x)$ is proportional to that
of $h(x)$ (which is $h_3$), through (\ref{constantapp}).

\,\,\,
\centerline{\bf Spin current and stress tensor}

Let us pose and understand the physical meaning of the modes we consider above.
%
The spin connection can be written with the metric, or the vielbein as eq.~(\ref{spin_con1}). 
This means that the spin current, which is dual to the spin connection, should be 
associated with the stress tensor, which is dual to the metric. 
Therefore we have to evaluate the spin current by taking into account its 
relation to the stress tensor.
In other words, the spin current can be determined by comparing
the coefficients appearing in the following relation,
\begin{equation}
 J^\mu_{~\hat{a}\hat{b}} \delta \omega_\mu^{~\hat{a}\hat{b}} = 
  T^{\rho\sigma}\delta \gamma_{\rho\sigma}
  =\delta {\cal L} \, .
  \label{varS}
\end{equation}
Here ${\cal L}$ 
is the Lagrangian of the quantum field theory in the boundary $2+1$ dimensions.  
Note that these metric and spin connection are defined on the boundary, therefore 
all the indices run without the radial direction. 
We have omitted the volume factor $\sqrt{-\gamma}$ for simplicity.

To obtain an explicit relation between the spin current and the stress
tensor, we first need to choose a local Lorentz frame. Any spin current is dependent on
the choice of the frame.  The boundary metric is
\bea
g_{tt}=-1\,,\,
g_{tx}= g_{xy} = -a \,,  \,
g_{xx}=1-a^2\,,\,
g_{yy}=1\,.
\eea
These are given by subtracting the scale factor $r$ of the bulk metric
in the limit $r\to\infty$.
A natural choice of the local Lorentz frame for the background 
vielbein consistent with this metric is given by \footnote{
When $a=0$, the vielbein is simply a unit matrix. The ``boost'' $t \rightarrow t + ax$
in the target space changes only the target space index $\mu$, resulting in
this form of the vielbein.
}
\bea
e_t^{\,\,\,\hat t} = 1  \,, \quad e_x^{\,\,\, \hat t} = a  \,,
\quad e_x^{\,\,\, \hat x} = 1 \, , \quad e_y^{\,\,\, \hat y} = 1 \, .
\eea
We turned on the AC fluctuation of the metric given by equation (\ref{ACflucepsilon}) and (\ref{ACflucepsilontwo}), and the most generic vielbein fluctuations consistent with 
 (\ref{ACflucepsilon}) and (\ref{ACflucepsilontwo}) is a set $\{e_t^{\,\,\, \hat y}, e_x^{\,\,\, \hat y},
 e_y^{\,\,\, \hat t}, e_y^{\,\,\, \hat x} \}$, which satisfies the two following relations
\bea
e_t^{\,\,\, \hat y} - e_y^{\,\,\, \hat t} & =& \epsilon \, e^{- i \omega t + i k_x x + i k_y y} f_0 \, , 
\label{rel-eg-1}\\
e_x^{\,\,\, \hat y} + e_y^{\,\,\, \hat x} - a e_y^{\,\,\, \hat t} &=&
\epsilon \, e^{- i \omega t + i k_x x + i k_y y} h_0 \, . 
\label{rel-eg-2}
\eea
coming from the constraint $\gamma_{\mu\nu} = e_{\mu}^{\,\,\, \hat a}  e_{\nu}^{\,\,\, \hat b} \eta_{\hat a \hat b}$.  
Here we used Fourier modes as $\sim e^{- i \omega t + i k_x x + i k_y y}$, and $(\omega, k_x,k_y)$ is the frequency/momentum for the fluctuations.
The other components of the vielbein are consistently put to zero in our case. 

With this at hand, all nontrivial components of the spin connection are
\bea
\delta \omega_t^{\,\,\, \hat t \hat y} &=& i \omega e_y^{\,\,\, \hat t} \, , 
\nonumber\\
\delta \omega_t^{\,\,\, \hat x \hat y} &=& 
- \frac{i}{2} k_x (e_t^{\,\,\, \hat y} - e_y^{\,\,\, \hat t})
+ \frac{i}{2} \omega (e_y^{\,\,\, \hat x} - e_x^{\,\,\, \hat y} 
+ a e_y^{\,\,\, \hat t})\, ,
\nonumber\\
\delta \omega_x^{\,\,\, \hat t \hat y} &=& 
- \frac{i}{2} k_x (e_t^{\,\,\, \hat y} + e_y^{\,\,\, \hat t})
+ \frac{i}{2} \omega (-e_y^{\,\,\, \hat x} - e_x^{\,\,\, \hat y} 
+ a e_y^{\,\,\, \hat t})\, ,
\nonumber\\
\delta \omega_x^{\,\,\, \hat x \hat y} &= &-
\frac{i}{2} k_x (2e_y^{\,\,\,\hat x} + ae_t^{\,\,\, \hat y} -a e_y^{\,\,\, \hat t})
\nonumber\\
& & + \frac{i}{2} a \omega (-e_y^{\,\,\, \hat x} - e_x^{\,\,\, \hat y} 
+ a e_y^{\,\,\, \hat t})\, ,
\nonumber\\
\delta \omega_y^{\,\,\, \hat t \hat x} &=& - 
\frac{i}{2} k_x (e_t^{\,\,\, \hat y} - e_y^{\,\,\, \hat t})
+ \frac{i}{2} \omega (-e_y^{\,\,\, \hat x} - e_x^{\,\,\, \hat y} 
+ a e_y^{\,\,\, \hat t})\, ,
\nonumber\\
\delta \omega_y^{\,\,\, \hat t \hat y} &=& 
- i k_y e_t^{\,\,\, \hat y}\, ,
\nonumber\\
\delta \omega_y^{\,\,\, \hat x \hat y} &=& - 
ik_y (a e_t^{\,\,\, \hat y} - e_x^{\,\,\, \hat y}) \, .
\label{spinconnections}
\eea
Keeping the two relations (\ref{rel-eg-1}) and (\ref{rel-eg-2}) satisfied, we can make a gauge choice
of the local Lorentz frame, $e_y^{\,\,\, \hat t} = e_y^{\,\,\, \hat x} = 0$, and restrict ourselves to
homogeneous fluctuation, $k_x=k_y=0$. In this local Lorentz frame, the above spin connections are simplified, and 
all the nonzero components are
\bea
\label{omegaxxyhzero}
&& \delta \omega_x^{\,\,\, \hat x \hat y} =
- \frac{i a \omega}{2}  \epsilon \, e^{- i \omega t } h_0 \,, \\
&& \delta \omega_t^{\,\,\, \hat x \hat y} 
= \delta \omega_x^{\,\,\, \hat t \hat y} = \delta \omega_y^{\,\,\, \hat t \hat x} =  
- \frac{i \omega }{2}  \epsilon \, e^{- i \omega t } h_0 \, ,
\label{omegatxyhzero}
\eea

Since $h_0$ is the constant mode of the boundary metric $g_{xy}$, it is a source for the
boundary stress tensor $T^{xy}$, therefore we obtain the spin current coupled to the spin connection from this expression as
\begin{eqnarray}
 &&
 J_x^{~\hat{x}\hat{y}} = - \frac{1}{a} \frac{1}{2 i \omega} T^{xy} \, , 
 \label{allspinconnectioncomponents}
\\
 &&
 J_t^{~\hat{x}\hat{y}} 
 = J_x^{~\hat{t}\hat{y}} 
 = J_y^{~\hat{t}\hat{x}} 
 = - \frac{1}{2 i \omega} T^{xy} \, .
  \label{allspinconnectioncomponentstwo}
\end{eqnarray} 
All the other components, other than each anti-symmetric partner  
$J_\mu^{\,\,\, \hat b \hat a} = - J_\mu^{\,\,\, \hat a \hat b}$, are zero.  
These combined with (\ref{omegaxxyhzero}) and (\ref{omegatxyhzero}) clearly satisfy (\ref{varS}). 
$ J_x^{\hat{x}\hat{y}}$ is the spin current along $x$ direction, and $ J_t^{~\hat{x}\hat{y}} 
 (= J_x^{~\hat{t}\hat{y}}  = J_y^{~\hat{t}\hat{x}} ) $ is the temporal component of the spin current, 
 corresponding to the spin density.

Here we have employed a choice of the local Lorentz frame 
$e_y^{\,\,\, \hat t} = e_y^{\,\,\, \hat x} = 0$. However other local Lorentz frame choices are also possible. 
Actually, for a certain other choice of the local Lorentz frame, one can show that the spin current
determined in this way is equivalent to a popular definition of the 
angular momentum current $M$ made by the stress-energy tensor, 
\bea
M_{\,\,\, \nu \lambda}^{\mu} \equiv
x_\nu T^{\mu}_{\,\,\, \lambda}- x_\lambda T^{\mu}_{\,\,\, \nu} \,.
\label{Mdef}
\eea
Due to this relation, for example, we can obtain the normalizable and
non-normalizable modes for the spin connection from those for the
metric.
Note that this current is with the target spacetime indices, so in order for this to
be equivalent to our spin current $J$, a certain local Lorentz frame should be appropriately
chosen.

To check this explicitly, we consider our case of nonzero $T^{ty}$ and $T^{xy}$.
We consider $a=0$ for simplicity. From the
definition (\ref{Mdef}), one obtains
\bea
&& M^{t}_{\,\,\, ty} = - t T^{ty}\, , \quad M^{t}_{\,\,\, xy} = x T^{ty}, \quad 
M^{x}_{\,\,\, ty} = - t T^{xy}\, , \nonumber \\
&& M^{x}_{\,\,\, xy} = x T^{xy}\, , \quad 
M^{y}_{\,\,\, tx} = x T^{ty} - t T^{xy}\, , \quad \nonumber \\
&& M^{y}_{\,\,\, ty} = y T^{ty}\, , \quad 
M^{y}_{\,\,\, xy} = -y T^{xy}\, . \quad 
\label{exM}
\eea
One can show that 
all of these 
are consistent with the spin connections (\ref{spinconnections})  
only when we choose a local Lorentz frame at which
\bea
e_t^{\,\,\, \hat y} = -e_y^{\,\,\, \hat t}\, , 
\quad
e_x^{\,\,\, \hat y} = e_y^{\,\,\, \hat x}
\eea
are satisfied. 
To see this, in this case, (\ref{rel-eg-1}), (\ref{rel-eg-2}) become 
\bea
 e_t^{\,\,\, \hat y} = - e_y^{\,\,\, \hat t}  =\frac{1}{2} \epsilon \, e^{- i \omega t + i k_x x + i k_y y} f_0 = \frac{1}{2} \delta g_{ty} \,, &&  \\
 e_x^{\,\,\, \hat y}= e_y^{\,\,\, \hat x} = \frac{1}{2} \epsilon \, e^{- i \omega t + i k_x x + i k_y y} h_0 
= \frac{1}{2} \delta g_{xy} \,, && 
\eea
and (\ref{spinconnections}) becomes
\bea
&& \delta \omega_t^{\,\,\, \hat t \hat y} 
=  \frac{1}{2} \partial_t \delta g_{ty} \, , \quad
\delta \omega_t^{\,\,\, \hat x \hat y} 
=   - \frac{1}{2} \partial_x \delta g_{ty} \,, \nonumber \\
&& \delta \omega_x^{\,\,\, \hat t \hat y} 
= \frac{1}{2} \partial_t \delta g_{xy}  \,, \quad 
\delta \omega_x^{\,\,\, \hat x \hat y} 
= -\frac{1}{2}  \partial_x \delta g_{xy} \,, \nonumber \\
&& \delta \omega_y^{\,\,\, \hat t \hat x}  
= - \frac{1}{2} \partial_x \delta g_{ty} + \frac{1}{2} \partial_t \delta g_{xy}\,, \nonumber \\
&& \delta \omega_y^{\,\,\, \hat t \hat y} 
=  - \frac{1}{2} \partial_y \delta g_{ty} \,, \quad 
\delta \omega_y^{\,\,\, \hat x \hat y} 
= \frac{1}{2} \partial_y \delta g_{xy}\,.
\label{spinconnections_simple}
\eea
Therefore the angular momentum current $M^\mu_{\, \, \, \nu \lambda}$ given by (\ref{exM}) satisfies our previous anticipation (\ref{varS}) with the spin connection (\ref{spinconnections_simple})
 via a partial integration.

The freedom for the local Lorentz frame choice 
corresponds to
the freedom for the local choice of the 
the axes to define the rotation for the angular momentum. 
Note that in any choice of the local Lorentz frame for the vielbein fluctuations, interestingly,
the expression of the most important spin connection (\ref{omegaxxyhzero}) is universal, therefore so 
is (\ref{allspinconnectioncomponents}).

\,\,\,
\centerline{\bf Transport coefficients}

$h_3$ is proportional to
the spin current $J_x^{\,\,\, \hat{z}} = 2 J_x^{\,\,\, \hat x \hat y}$.
$h_0$ is proportional to the spin 
gradient along the $x$ direction $\nabla_x \mu^{\hat{z}}$, because 
$\nabla_x \mu^{\hat{z}} = 2 \nabla_x \omega_{t ~{\rm (NN})}^{\,\,\, \hat x \hat y}$ is 
gauge equivalent to $-2 \nabla_t \omega_{x ~{\rm (NN})}^{\,\,\, \hat x \hat y} = 2 i \omega \, \omega_{x ~{\rm (NN})}^{\,\,\, \hat{x}\hat{y}}$.

$f_0$ corresponds to the thermal gradient along the $y$ direction 
due to the relation $i\omega \delta g_{ty}^{\rm (NN)} =  r^2 \nabla_{y} T/T$ 
\footnote{This can be derived by scaling time in the unit of temperature 
as $g_{tt}^{\rm boundary} = -1/T^2$, 
and by using a gauge transformation, which transform $\nabla_x g_{tt}^{\rm bulk}$ to $- 2 \nabla_t g_{tx}^{\rm bulk}$. 
The extra $r^2$ is because of $g_{\mu\nu}^{\rm bulk} = r^2 g_{\mu\nu}^{\rm boundary}$.}. 
$f_3$ corresponds to a {\it thermal current} along the $y$ direction, since $\delta g_{ty}^{\rm (N)}$ is dual to the stress-tensor, $\delta g_{ty}^{\rm (N)} = T_{ty}$.

From these, we can evaluate the spin transport coefficient $\alpha$, 
and the thermal spin Hall conductivity $\kappa_{s{\rm H}}$,
defined as  
\bea
 J_x^{\,\,\, \hat{z}} = - \alpha \nabla_x \mu^{\hat{z}}, \qquad
 J_x^{\,\,\, \hat{z}} = - \kappa_{s{\rm H}} \nabla_y T \,.
\eea
Using holography, these coefficients are represented by normalizable ${\rm (N)}$ and
non-normalizable modes ${\rm (NN)}$ as 
\bea
\alpha &=& - \frac{J_x^{\,\,\, \hat{x}\hat{y}}}{i\omega \delta
\omega_{x~{\rm (NN)}}^{\,\,\, \hat{x}\hat{y}}} =  \frac{h_3}{ia^2 \omega^3 h_0} \,, \\
\kappa_{s{\rm H}}
&=& - \frac{2J_x^{\,\,\, \hat{x}\hat{y}}}{ {i\omega T\delta g_{ty}^{\rm (NN)}}/{r^2}} 
=  -\frac{  h_3}{a \omega^2 T f_0   }    \,.
\eea
As we have seen, the ratio $h_3/h_0$ and $h_3/f_0$ are functions of 
only $\omega/T$, in equations (\ref{eqx}),  (\ref{horizonx}) and (\ref{constantapp}). 
We obtain these, by solving the bulk equation and imposing the in-going boundary 
condition at the horizon, and 
the radial $r$ dependence of the bulk equation is reflected as $\omega/r_0$ dependence in 
the boundary viewpoint. 

Actually the sources $(h_0, f_0)$ and the expectation values $(h_3, f_3)$ 
are related by a 2 by 2 matrix, and the coefficients $\alpha$ and $\kappa_{\rm sH}$ are just upper 
2 elements of this 2 by 2 matrix.  
However, as we have seen, in our system it follows $f_3 = (a - 1/a)^{-1} h_3$ due to the relation (\ref{Einsry}), 
where $f_3$ is the normalizable mode coefficient for $f(r)$, just as $h_3$ in the equation (\ref{h3def}). 
Therefore, the ratio $f_3/f_0$ and the ratio $f_3/h_0$ 
are essentially the same as $h_3/f_0$ and the ratio $h_3/h_0$.  

\begin{figure}[t]
 \begin{center}
  \includegraphics[height=15em]{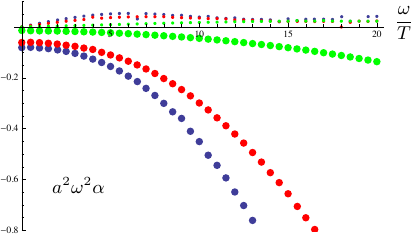}
  \\ \vspace{2em}
  \includegraphics[height=15em]{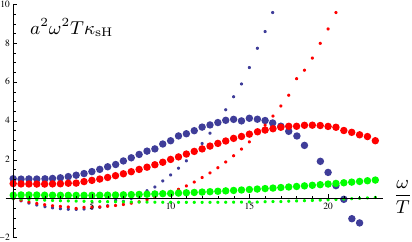}
 \end{center}
 \caption{Top: the spin transport coefficient $\alpha$ as a function of the
 frequency over the temperature, $\omega/T$. 
 Large dots are the real part Re$[\alpha]$, and small 
 dots are the imaginary part Im$[\alpha]$. Blue, red and green correspond to
 $a =0.03, 0.5$ and $0.9$, respectively.
  Bottom: the thermal spin Hall conductivity $\kappa_{s{\rm H}}$ as a
 function of $\omega/T$. 
 }
 \label{figs1}
\end{figure}

We have evaluated these transport coefficients by a numerical method for solving the differential 
equation (\ref{eqx}). By varying the frequency $\omega$, 
we find the AC conductivities as shown in Fig.~\ref{figs1} \footnote{
One might wonder if Onsager's reciprocal relation holds in this case. 
Since we have two thermodynamical quantities represented by $f(r)$ and $h(r)$ in the holographic language, it is natural to argue the reciprocal
relation. There are two points concerning the relation. First, since we have introduced the background
AC external source for $g_{ty}$ and also the background metric $g_{tx}$, it is expected that we 
explicitly break the time-reversal symmetry. So there is no good reason for the reciprocal relation to hold in our case. Second, as we have noticed, once $h(r)$ is given, then $h(r)$ is completely determined. So,
we cannot turn on the external source for $g_{xy}$ and $g_{ty}$
independently. This means that the Onsager reciprocal relation is not
directly measured by our external sources.}.

For the numerical simulations, we have worked in the unit $T=1$ 
and chosen $a=0.03, a=0.5$ and $a=0.9$ for simplicity.
The top figure of Fig.~\ref{figs1} is the spin transport coefficient
$\alpha$. This is the coefficient on the spin current
$J_x^{\,\,\, \hat{z}}$ as a response to the AC external gradient of the spin
chemical potential $\mu^{\hat{z}}$.
The bottom figure of Fig.~\ref{figs1} is the thermal spin Hall conductivity $\kappa_{\rm sH}$. 
In both figures, the transport coefficients are multiplied by $a^2 \omega^2$ to show
the $\omega/T$ dependence clearly.
From the figures, we find that the imaginary parts $\times \, \omega^2$ vanish 
linearly at $\omega=0$, so around the origin the imaginary parts behave as $1/\omega$. 
This means that in the real parts there exists a Drude peak proportional to $\delta(\omega)$ 
often observed in superconducting/metal phases.
We also see specific behavior of the thermal spin Hall conductivity,
changing the sign of the transport coefficient as the frequency gets larger.
It is quite interesting to observe such frequency dependence by experimental or
other theoretical setups.

%

%

\,\,\,
\centerline{\bf On the spin current definition}
We have evaluated the spin current following the relation (\ref{varS}). 
However, (\ref{varS}) is not necessarily the same as our definition of the 
spin current (\ref{spin_curr1}). We will now discuss that the spin current evaluated by the 
definition (\ref{spin_curr1}) yields zero value, using the action
\eqref{EH_action}
\footnote{Here we do not take into account the cosmological counter term for
simplicity.}.
This is the reason why we need to relate the spin current to the stress tensor 
as (\ref{varS}) which we have used in this paper. 


To obtain the spin current following the definition
(\ref{spin_curr1}) in holography, note that (\ref{spin_curr1}) means
that we have to differentiate the action (\ref{EH_action}) with the {\em
boundary spin connection}, which is defined by the spin connections
along the boundary directions. 
The contribution coming from a variation of the bulk action
(\ref{bulk_theory}) by the boundary spin connection, vanishes by using
the bulk equations of motion. 
Thus the contribution to the spin current comes from a variation of the
boundary action (\ref{boundary_theory}) 
only. 
However we will see that this contribution also vanishes.

Whenever we take a variation, we have to fix all the other quantities. 
In this case, we regard each of the boundary spin connection components
as an independent degree of freedom, and then, we take a variation of
the action by that, while keeping all the other quantities, which
include metric, fixed.
In this formulation, each spin connection component is an independent
degree of freedom from the metric; The independent degrees of freedom
are metric and spin connection.
In fact, we can formulate general relativity in such a way, by  {\it
i.e.,} so-called Palatini formulation of gravity. 
However this procedure turns out to give a vanishing spin current.

To see this, let us conduct a variation of the boundary action (\ref{boundary_theory}) by the boundary spin connection. 
The extrinsic curvature $\Theta$ is written with the
normal vector 
$n_\mu$ as  
$\Theta = - \gamma^{\mu \nu} \nabla_{\mu} n_\nu $.  
%
In the Palatini formalism, the boundary metric $\gamma^{\mu\nu}$ and the boundary spin connection are 
independent, 
therefore, the contribution form the boundary action variation yields 
\bea
 J^\mu_{\,\,\, \hat{a}\hat{b}}  = - 2 \gamma^{\rho\sigma} \frac{\delta \Gamma^\eta_{\xi\lambda}}{\delta
  \omega_\mu^{\,\,\, \hat{a}\hat{b}}} \frac{\delta (\nabla_\rho n_\sigma)}{\delta \Gamma^\eta_{\xi\lambda}} 
= - 2 e^\mu_{\,\,\,\hat{a}}e^\nu_{\,\,\,\hat{b}}n_\nu \,.
\label{Jw}
\eea
Since $n_{\nu}\not=0$ only when $\nu=r$ and $e^r_{\,\,\,\hat{b}}\neq 0$ 
only when $\hat{b}=\hat{r}$, there is no spin
current on the boundary. 
This shows that the spin current evaluated by the Palatini formalism vanishes 
\footnote{In this evaluation, we have used the bulk equation of motion (\ref{spin_con1}) to 
define $\Gamma^\eta_{\xi\lambda} (e_\mu^{\,\,\,\hat{a}}, \omega_\mu^{\,\,\, \hat{a}\hat{b}})$. 
Instead, if we regard the extrinsic curvature $\Theta$ as being solely
written by the vielbein, the variation (\ref{Jw}) vanishes in the
Palatini formalism.}. 
To obtain a non-vanishing spin current, 
we should {\it not} regard the metric and the spin connection as independent degrees of freedom.
We need to modify our definition of the spin current (\ref{spin_curr1}) slightly.

Therefore in this paper, 
we do not regard the spin connection as an independent
variable, but associate it with the metric.
This further implies that our spin current, which is dual to the spin connection, should be 
associated with the stress tensor, which is dual to the metric. 
In the Palatini formalism, the relation (\ref{spin_con1}) comes from the equation of motion for the spin connection. 
Therefore we have evaluated the spin current by taking into account its 
relation to the stress tensor as (\ref{varS}) in this paper.

\,\,\,
\centerline{\bf Discussions : Spin vs angular momentum}

In this paper, we have investigated the spin transport phenomena from the view
point of gauge/gravity correspondence. 
We have introduced the proper definition of the spin current, 
as a conserved N\"other's current, which couples naturally to the 
spin connection.

We have analyzed the AdS Schwarzschild black brane geometry as a simple
example to demonstrate how to study the spin transport in the context of
the holography.
We have calculated the spin transport coefficient $\alpha$ 
and the thermal spin Hall conductivity $\kappa_{\rm sH}$ 
by studying the fluctuations of the metric components. 
We have obtained the corresponding transport coefficient from the
non-normalizable and normalizable modes propagating in the bulk gravity.

\begin{figure}[t]
 \begin{center}
  \includegraphics[width=20em]{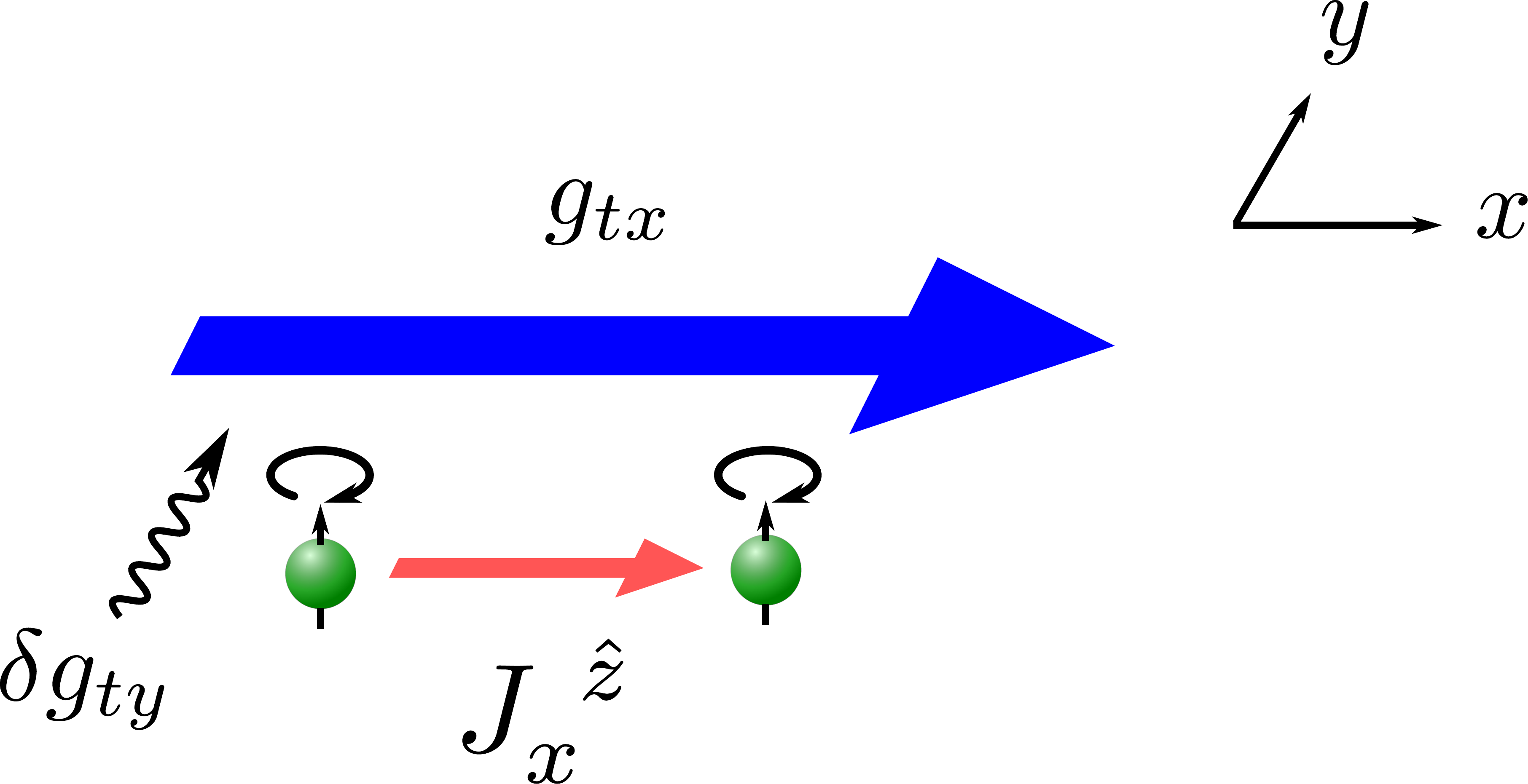}
 \end{center}
 \caption{When the off-diagonal background metric $g_{tx}$,
 namely a constant energy flow in $x$-direction, is turned on,
 the angular momentum current as a spin current $J_x^{\,\,\, \hat{z}}$ is
 induced by applying the fluctuation $\delta g_{ty}$.}
 \label{spin_current2}
\end{figure}

Let us comment on a physical meaning of the holographic analysis done in
this paper.
We have seen that the off-diagonal metric component for the background, {\it i.e.}, $g_{tx} (= g_{xy})$, is required
for giving the spin current.
Note that if there is such a component in the background geometry, that leads to a
constant energy flow coupled to $g_{tx}$. 
By applying the fluctuation $\delta g_{ty}$ in addition to the
background flow, we should have an angular momentum 
current in $x$-direction as shown in
Fig.~\ref{spin_current2}.
It seems that our spin current almost corresponds to the orbital part
of angular momentum.

However, at least from the relativistic theoretical viewpoint, 
we cannot split the total angular momentum into contributions from orbital and 
intrinsic spin; 
Spin is originally defined in the non-relativistic system, where the Lorentz invariance is 
broken and we should treat space and time separately. 
Since in this paper we have considered the ``total angular momentum'' current 
defined in {\it relativistic} field theory, 
in order to really discuss the ``spin''-current, we need to take an appropriate 
non-relativistic limit of our system. 
Only after taking that, we can extrapolate the spin contribution 
from the total angular momentum current, 
and we can discuss if the 
orbital contribution gives only a sub-leading contribution or not. 

The non-relativistic limit of relativistic conformal field theories
is obtained by taking the {\it discreet light-cone quantization} (DLCQ). 
This limit reduces the boundary metric from $AdS$ into the following form \cite{Son:2008ye,Balasubramanian:2008dm,Herzog:2008wg,Maldacena:2008wh,Adams:2008wt} 
\bea
\label{DLCQmetric}
ds^2 = - {r^{2z}}  {(dx^{+})^2}+ \frac{dr^2}{r^2}+  2  r^2 { d{x^+} d{x^-}} + r^2 {d\vec{x}^2}  \,,
\eea
where $x^{+}$ is the light front time, $r$ is the holographic radial
direction as before. 
$x^{-}$ is a new direction associated with the boost direction and we compactly $x^- \sim x^- + R$, and has an interpretation as ``dual'' to the conserved particle number since $P_{-}$ is quantized as $N/R$, where $N$ is particle number. $z$ is called ``dynamical exponent'' and represents the 
difference of the scaling between time $x^+$ and spatial coordinate $\vec{x}$. 
 
For example, starting from a boundary theory which is 3+1 dimensional, we can obtain a 
2+1-dimensional non-relativistic theory where we can identify $x^+ = t + x^3$ and $x^- = t - x^3$. 
This metric possesses  the Schr\"odinger symmetry for the $z=2$ case. 

Taking this DLCQ limit, or simply replacing the boundary metric from AdS into the above, is not enough 
for extracting the spin information, 
since spin is not a conserved quantity by itself even here, and only the 
total angular momentum is a conserved one. 
To eliminate the contribution of the orbital angular momentum, 
it is best to consider a setting where 
the momentum of the particle is suppressed, namely an insulator. 
The insulator is realized as a system which has an energy gap. 
The energy gap is reflected in a holographic setting in the bulk as a system which has 
an IR cut-off, 
like the confinement in holographic QCD. The hard wall model is the simplest setting 
to realize the mass gap and therefore this would lead one to a system 
which has an asymptotic metric as (\ref{DLCQmetric}) and has an IR cut-off. 
Such a bulk set-up is good for us to study the spin-transport phenomena 
and 
it is interesting to see how the orbital 
contribution and the real `spin' contribution to our total spin current, 
after taking the non-relativistic limit.

In this paper, we considered only the spin-current induction by the spin-current potential and also 
thermo-potential, but not the one induced by an electric field. In real experiments, it is more often to consider the spin current induced by some external electric field, so this forces us to consider a bulk action coupled to the 
electromagnetic field. 
Adding impurity effects \cite{Harrison:2011fs,Hartnoll:2008hs,Adams:2011rj,Hashimoto:2012pb}  is also important. 
We hope to return to these analyses in the near future. 

\,
\centerline{\bf Acknowledgements}
N.I. would like to thank RIKEN Mathematical Physics Laboratory for kind hospitality where this project started. 
The research of K.H. is supported
in part by JSPS Grants-in-Aid for Scientific Research No.~23105716, 23654096, 22340069.
The research of T.K. is supported in part by Grant-in-Aid for JSPS Fellows~(No.~23-593).



\begin{thebibliography}{30}%
\makeatletter
\providecommand \@ifxundefined [1]{%
 \@ifx{#1\undefined}
}%
\providecommand \@ifnum [1]{%
 \ifnum #1\expandafter \@firstoftwo
 \else \expandafter \@secondoftwo
 \fi
}%
\providecommand \@ifx [1]{%
 \ifx #1\expandafter \@firstoftwo
 \else \expandafter \@secondoftwo
 \fi
}%
\providecommand \natexlab [1]{#1}%
\providecommand \enquote  [1]{``#1''}%
\providecommand \bibnamefont  [1]{#1}%
\providecommand \bibfnamefont [1]{#1}%
\providecommand \citenamefont [1]{#1}%
\providecommand \href@noop [0]{\@secondoftwo}%
\providecommand \href [0]{\begingroup \@sanitize@url \@href}%
\providecommand \@href[1]{\@@startlink{#1}\@@href}%
\providecommand \@@href[1]{\endgroup#1\@@endlink}%
\providecommand \@sanitize@url [0]{\catcode `\\12\catcode `\$12\catcode
  `\&12\catcode `\#12\catcode `\^12\catcode `\_12\catcode `\%12\relax}%
\providecommand \@@startlink[1]{}%
\providecommand \@@endlink[0]{}%
\providecommand \url  [0]{\begingroup\@sanitize@url \@url }%
\providecommand \@url [1]{\endgroup\@href {#1}{\urlprefix }}%
\providecommand \urlprefix  [0]{URL }%
\providecommand \Eprint [0]{\href }%
\providecommand \doibase [0]{http://dx.doi.org/}%
\providecommand \selectlanguage [0]{\@gobble}%
\providecommand \bibinfo  [0]{\@secondoftwo}%
\providecommand \bibfield  [0]{\@secondoftwo}%
\providecommand \translation [1]{[#1]}%
\providecommand \BibitemOpen [0]{}%
\providecommand \bibitemStop [0]{}%
\providecommand \bibitemNoStop [0]{.\EOS\space}%
\providecommand \EOS [0]{\spacefactor3000\relax}%
\providecommand \BibitemShut  [1]{\csname bibitem#1\endcsname}%
\let\auto@bib@innerbib\@empty
\bibitem [{\citenamefont {Maekawa}(2006)}]{Maekawa:2006OUP}%
  \BibitemOpen
  \bibinfo {editor} {\bibfnamefont {S.}~\bibnamefont {Maekawa}},\ ed.,\ \href
  {\doibase 10.1093/acprof:oso/9780198568216.001.0001} {\emph {\bibinfo {title}
  {{Concepts in Spin Electronics}}}}\ (\bibinfo  {publisher} {Oxford Univ.
  Press},\ \bibinfo {year} {2006})\BibitemShut {NoStop}%
\bibitem [{\citenamefont {\v{Z}uti\'c}\ and\ \citenamefont
  {Dery}(2011)}]{Zutic:2011NM}%
  \BibitemOpen
  \bibfield  {author} {\bibinfo {author} {\bibfnamefont {I.}~\bibnamefont
  {\v{Z}uti\'c}}\ and\ \bibinfo {author} {\bibfnamefont {H.}~\bibnamefont
  {Dery}},\ }\href {\doibase 10.1038/nmat3097} {\bibfield  {journal} {\bibinfo
  {journal} {Nature Materials}\ }\textbf {\bibinfo {volume} {10}},\ \bibinfo
  {pages} {647} (\bibinfo {year} {2011})}\BibitemShut {NoStop}%
\bibitem [{\citenamefont {Maldacena}(1998)}]{Maldacena:1997re}%
  \BibitemOpen
  \bibfield  {author} {\bibinfo {author} {\bibfnamefont {J.~M.}\ \bibnamefont
  {Maldacena}},\ }\href {\doibase 10.1023/A:1026654312961} {\bibfield
  {journal} {\bibinfo  {journal} {Adv. Theor. Math. Phys.}\ }\textbf {\bibinfo
  {volume} {2}},\ \bibinfo {pages} {231} (\bibinfo {year} {1998})},\ \Eprint
  {http://arxiv.org/abs/hep-th/9711200} {arXiv:hep-th/9711200} \BibitemShut
  {NoStop}%
\bibitem [{\citenamefont {Gubser}\ \emph {et~al.}(1998)\citenamefont {Gubser},
  \citenamefont {Klebanov},\ and\ \citenamefont {Polyakov}}]{Gubser:1998bc}%
  \BibitemOpen
  \bibfield  {author} {\bibinfo {author} {\bibfnamefont {S.}~\bibnamefont
  {Gubser}}, \bibinfo {author} {\bibfnamefont {I.~R.}\ \bibnamefont
  {Klebanov}}, \ and\ \bibinfo {author} {\bibfnamefont {A.~M.}\ \bibnamefont
  {Polyakov}},\ }\href {\doibase 10.1016/S0370-2693(98)00377-3} {\bibfield
  {journal} {\bibinfo  {journal} {Phys. Lett.}\ }\textbf {\bibinfo {volume}
  {B428}},\ \bibinfo {pages} {105} (\bibinfo {year} {1998})},\ \Eprint
  {http://arxiv.org/abs/hep-th/9802109} {arXiv:hep-th/9802109 [hep-th]}
  \BibitemShut {NoStop}%
\bibitem [{\citenamefont {Witten}(1998)}]{Witten:1998qj}%
  \BibitemOpen
  \bibfield  {author} {\bibinfo {author} {\bibfnamefont {E.}~\bibnamefont
  {Witten}},\ }\href@noop {} {\bibfield  {journal} {\bibinfo  {journal} {Adv.
  Theor. Math. Phys.}\ }\textbf {\bibinfo {volume} {2}},\ \bibinfo {pages}
  {253} (\bibinfo {year} {1998})},\ \Eprint
  {http://arxiv.org/abs/hep-th/9802150} {arXiv:hep-th/9802150 [hep-th]}
  \BibitemShut {NoStop}%
\bibitem [{\citenamefont {Benini}\ \emph {et~al.}(2010)\citenamefont {Benini},
  \citenamefont {Herzog}, \citenamefont {Rahman},\ and\ \citenamefont
  {Yarom}}]{Benini:2010pr}%
  \BibitemOpen
  \bibfield  {author} {\bibinfo {author} {\bibfnamefont {F.}~\bibnamefont
  {Benini}}, \bibinfo {author} {\bibfnamefont {C.~P.}\ \bibnamefont {Herzog}},
  \bibinfo {author} {\bibfnamefont {R.}~\bibnamefont {Rahman}}, \ and\ \bibinfo
  {author} {\bibfnamefont {A.}~\bibnamefont {Yarom}},\ }\href {\doibase
  10.1007/JHEP11(2010)137} {\bibfield  {journal} {\bibinfo  {journal} {JHEP}\
  }\textbf {\bibinfo {volume} {1011}},\ \bibinfo {pages} {137} (\bibinfo {year}
  {2010})},\ \Eprint {http://arxiv.org/abs/1007.1981} {arXiv:1007.1981
  [hep-th]} \BibitemShut {NoStop}%
\bibitem [{\citenamefont {Harrison}\ \emph {et~al.}(2012)\citenamefont
  {Harrison}, \citenamefont {Kachru},\ and\ \citenamefont
  {Torroba}}]{Harrison:2011fs}%
  \BibitemOpen
  \bibfield  {author} {\bibinfo {author} {\bibfnamefont {S.}~\bibnamefont
  {Harrison}}, \bibinfo {author} {\bibfnamefont {S.}~\bibnamefont {Kachru}}, \
  and\ \bibinfo {author} {\bibfnamefont {G.}~\bibnamefont {Torroba}},\ }\href
  {\doibase 10.1088/0264-9381/29/19/194005} {\bibfield  {journal} {\bibinfo
  {journal} {Class. Quant. Grav.}\ }\textbf {\bibinfo {volume} {29}},\ \bibinfo
  {pages} {194005} (\bibinfo {year} {2012})},\ \Eprint
  {http://arxiv.org/abs/1110.5325} {arXiv:1110.5325 [hep-th]} \BibitemShut
  {NoStop}%
\bibitem [{\citenamefont {Bigazzi}\ \emph {et~al.}(2012)\citenamefont
  {Bigazzi}, \citenamefont {Cotrone}, \citenamefont {Musso}, \citenamefont
  {Fokeeva},\ and\ \citenamefont {Seminara}}]{Bigazzi:2011ak}%
  \BibitemOpen
  \bibfield  {author} {\bibinfo {author} {\bibfnamefont {F.}~\bibnamefont
  {Bigazzi}}, \bibinfo {author} {\bibfnamefont {A.~L.}\ \bibnamefont
  {Cotrone}}, \bibinfo {author} {\bibfnamefont {D.}~\bibnamefont {Musso}},
  \bibinfo {author} {\bibfnamefont {N.~P.}\ \bibnamefont {Fokeeva}}, \ and\
  \bibinfo {author} {\bibfnamefont {D.}~\bibnamefont {Seminara}},\ }\href
  {\doibase 10.1007/JHEP02(2012)078} {\bibfield  {journal} {\bibinfo  {journal}
  {JHEP}\ }\textbf {\bibinfo {volume} {1202}},\ \bibinfo {pages} {078}
  (\bibinfo {year} {2012})},\ \Eprint {http://arxiv.org/abs/1111.6601}
  {arXiv:1111.6601 [hep-th]} \BibitemShut {NoStop}%
\bibitem [{\citenamefont {Herzog}\ and\ \citenamefont
  {Ren}(2012)}]{Herzog:2012kx}%
  \BibitemOpen
  \bibfield  {author} {\bibinfo {author} {\bibfnamefont {C.~P.}\ \bibnamefont
  {Herzog}}\ and\ \bibinfo {author} {\bibfnamefont {J.}~\bibnamefont {Ren}},\
  }\href {\doibase 10.1007/JHEP06(2012)078} {\bibfield  {journal} {\bibinfo
  {journal} {JHEP}\ }\textbf {\bibinfo {volume} {1206}},\ \bibinfo {pages}
  {078} (\bibinfo {year} {2012})},\ \Eprint {http://arxiv.org/abs/1204.0518}
  {arXiv:1204.0518 [hep-th]} \BibitemShut {NoStop}%
\bibitem [{\citenamefont {Benincasa}\ and\ \citenamefont
  {Ramallo}(2012)}]{Benincasa:2012wu}%
  \BibitemOpen
  \bibfield  {author} {\bibinfo {author} {\bibfnamefont {P.}~\bibnamefont
  {Benincasa}}\ and\ \bibinfo {author} {\bibfnamefont {A.~V.}\ \bibnamefont
  {Ramallo}},\ }\href {\doibase 10.1007/JHEP06(2012)133} {\bibfield  {journal}
  {\bibinfo  {journal} {JHEP}\ }\textbf {\bibinfo {volume} {1206}},\ \bibinfo
  {pages} {133} (\bibinfo {year} {2012})},\ \Eprint
  {http://arxiv.org/abs/1204.6290} {arXiv:1204.6290 [hep-th]} \BibitemShut
  {NoStop}%
\bibitem [{\citenamefont {Alexandrov}\ and\ \citenamefont
  {Coleman}(2012)}]{PhysRevB.86.125145}%
  \BibitemOpen
  \bibfield  {author} {\bibinfo {author} {\bibfnamefont {V.}~\bibnamefont
  {Alexandrov}}\ and\ \bibinfo {author} {\bibfnamefont {P.}~\bibnamefont
  {Coleman}},\ }\href {\doibase 10.1103/PhysRevB.86.125145} {\bibfield
  {journal} {\bibinfo  {journal} {Phys. Rev.}\ }\textbf {\bibinfo {volume}
  {B86}},\ \bibinfo {pages} {125145} (\bibinfo {year} {2012})},\ \Eprint
  {http://arxiv.org/abs/1204.6310} {arXiv:1204.6310 [cond-mat.str-el]}
  \BibitemShut {NoStop}%
\bibitem [{\citenamefont {Luo}(2012)}]{Luo:2012am}%
  \BibitemOpen
  \bibfield  {author} {\bibinfo {author} {\bibfnamefont {M.}~\bibnamefont
  {Luo}},\ }\href@noop {} {\  (\bibinfo {year} {2012})},\ \Eprint
  {http://arxiv.org/abs/1205.3267} {arXiv:1205.3267 [hep-th]} \BibitemShut
  {NoStop}%
\bibitem [{\citenamefont {Hashimoto}\ and\ \citenamefont
  {Iizuka}(2012)}]{Hashimoto:2012pb}%
  \BibitemOpen
  \bibfield  {author} {\bibinfo {author} {\bibfnamefont {K.}~\bibnamefont
  {Hashimoto}}\ and\ \bibinfo {author} {\bibfnamefont {N.}~\bibnamefont
  {Iizuka}},\ }\href@noop {} {\  (\bibinfo {year} {2012})},\ \Eprint
  {http://arxiv.org/abs/1207.4643} {arXiv:1207.4643 [hep-th]} \BibitemShut
  {NoStop}%
\bibitem [{\citenamefont {Ishii}\ and\ \citenamefont
  {Sin}(2013)}]{Ishii:2012hw}%
  \BibitemOpen
  \bibfield  {author} {\bibinfo {author} {\bibfnamefont {T.}~\bibnamefont
  {Ishii}}\ and\ \bibinfo {author} {\bibfnamefont {S.-J.}\ \bibnamefont
  {Sin}},\ }\href {\doibase 10.1007/JHEP04(2013)128} {\bibfield  {journal}
  {\bibinfo  {journal} {JHEP}\ }\textbf {\bibinfo {volume} {1304}},\ \bibinfo
  {pages} {128} (\bibinfo {year} {2013})},\ \Eprint
  {http://arxiv.org/abs/1211.1798} {arXiv:1211.1798 [hep-th]} \BibitemShut
  {NoStop}%
\bibitem [{\citenamefont {Wen}\ and\ \citenamefont {Zee}(1992)}]{Wen:1992ej}%
  \BibitemOpen
  \bibfield  {author} {\bibinfo {author} {\bibfnamefont {X.~G.}\ \bibnamefont
  {Wen}}\ and\ \bibinfo {author} {\bibfnamefont {A.}~\bibnamefont {Zee}},\
  }\href {\doibase 10.1103/PhysRevLett.69.953} {\bibfield  {journal} {\bibinfo
  {journal} {Phys. Rev. Lett.}\ }\textbf {\bibinfo {volume} {69}},\ \bibinfo
  {pages} {953} (\bibinfo {year} {1992})}\BibitemShut {NoStop}%
\bibitem [{\citenamefont {Fr\"ohlich}\ and\ \citenamefont
  {Studer}(1993)}]{RevModPhys.65.733}%
  \BibitemOpen
  \bibfield  {author} {\bibinfo {author} {\bibfnamefont {J.}~\bibnamefont
  {Fr\"ohlich}}\ and\ \bibinfo {author} {\bibfnamefont {U.~M.}\ \bibnamefont
  {Studer}},\ }\href {\doibase 10.1103/RevModPhys.65.733} {\bibfield  {journal}
  {\bibinfo  {journal} {Rev. Mod. Phys.}\ }\textbf {\bibinfo {volume} {65}},\
  \bibinfo {pages} {733} (\bibinfo {year} {1993})}\BibitemShut {NoStop}%
\bibitem [{\citenamefont {Mineev}\ and\ \citenamefont
  {Volovik}(1992)}]{Mineev:JLTP1989}%
  \BibitemOpen
  \bibfield  {author} {\bibinfo {author} {\bibfnamefont {V.}~\bibnamefont
  {Mineev}}\ and\ \bibinfo {author} {\bibfnamefont {G.}~\bibnamefont
  {Volovik}},\ }\href {\doibase 10.1007/BF00683888} {\bibfield  {journal}
  {\bibinfo  {journal} {J. Low Temp. Phys.}\ }\textbf {\bibinfo {volume}
  {89}},\ \bibinfo {pages} {823} (\bibinfo {year} {1992})}\BibitemShut
  {NoStop}%
\bibitem [{\citenamefont {Goldhaber}(1989)}]{PhysRevLett.62.482}%
  \BibitemOpen
  \bibfield  {author} {\bibinfo {author} {\bibfnamefont {A.~S.}\ \bibnamefont
  {Goldhaber}},\ }\href {\doibase 10.1103/PhysRevLett.62.482} {\bibfield
  {journal} {\bibinfo  {journal} {Phys. Rev. Lett.}\ }\textbf {\bibinfo
  {volume} {62}},\ \bibinfo {pages} {482} (\bibinfo {year} {1989})}\BibitemShut
  {NoStop}%
\bibitem [{\citenamefont {Fr\"ohlich}\ and\ \citenamefont
  {Studer}(1992)}]{Frohlich:1992CMP}%
  \BibitemOpen
  \bibfield  {author} {\bibinfo {author} {\bibfnamefont {J.}~\bibnamefont
  {Fr\"ohlich}}\ and\ \bibinfo {author} {\bibfnamefont {U.}~\bibnamefont
  {Studer}},\ }\href {\doibase 10.1007/BF02096549} {\bibfield  {journal}
  {\bibinfo  {journal} {Commun. Math. Phys.}\ }\textbf {\bibinfo {volume}
  {148}},\ \bibinfo {pages} {553} (\bibinfo {year} {1992})}\BibitemShut
  {NoStop}%
\bibitem [{\citenamefont {Balasubramanian}\ and\ \citenamefont
  {Kraus}(1999)}]{Balasubramanian:1999re}%
  \BibitemOpen
  \bibfield  {author} {\bibinfo {author} {\bibfnamefont {V.}~\bibnamefont
  {Balasubramanian}}\ and\ \bibinfo {author} {\bibfnamefont {P.}~\bibnamefont
  {Kraus}},\ }\href {\doibase 10.1007/s002200050764} {\bibfield  {journal}
  {\bibinfo  {journal} {Commun. Math. Phys.}\ }\textbf {\bibinfo {volume}
  {208}},\ \bibinfo {pages} {413} (\bibinfo {year} {1999})},\ \Eprint
  {http://arxiv.org/abs/hep-th/9902121} {arXiv:hep-th/9902121 [hep-th]}
  \BibitemShut {NoStop}%
\bibitem [{\citenamefont {de~Haro}\ \emph {et~al.}(2001)\citenamefont
  {de~Haro}, \citenamefont {Solodukhin},\ and\ \citenamefont
  {Skenderis}}]{deHaro:2000xn}%
  \BibitemOpen
  \bibfield  {author} {\bibinfo {author} {\bibfnamefont {S.}~\bibnamefont
  {de~Haro}}, \bibinfo {author} {\bibfnamefont {S.~N.}\ \bibnamefont
  {Solodukhin}}, \ and\ \bibinfo {author} {\bibfnamefont {K.}~\bibnamefont
  {Skenderis}},\ }\href {\doibase 10.1007/s002200100381} {\bibfield  {journal}
  {\bibinfo  {journal} {Commun. Math. Phys.}\ }\textbf {\bibinfo {volume}
  {217}},\ \bibinfo {pages} {595} (\bibinfo {year} {2001})},\ \Eprint
  {http://arxiv.org/abs/hep-th/0002230} {arXiv:hep-th/0002230 [hep-th]}
  \BibitemShut {NoStop}%
\bibitem [{\citenamefont {Kovtun}\ \emph {et~al.}(2005)\citenamefont {Kovtun},
  \citenamefont {Son},\ and\ \citenamefont {Starinets}}]{Kovtun:2004de}%
  \BibitemOpen
  \bibfield  {author} {\bibinfo {author} {\bibfnamefont {P.~K.}\ \bibnamefont
  {Kovtun}}, \bibinfo {author} {\bibfnamefont {D.~T.}\ \bibnamefont {Son}}, \
  and\ \bibinfo {author} {\bibfnamefont {A.~O.}\ \bibnamefont {Starinets}},\
  }\href {\doibase 10.1103/PhysRevLett.94.111601} {\bibfield  {journal}
  {\bibinfo  {journal} {Phys. Rev. Lett.}\ }\textbf {\bibinfo {volume} {94}},\
  \bibinfo {pages} {111601} (\bibinfo {year} {2005})},\ \Eprint
  {http://arxiv.org/abs/hep-th/0405231} {arXiv:hep-th/0405231 [hep-th]}
  \BibitemShut {NoStop}%
\bibitem [{\citenamefont {Son}(2008)}]{Son:2008ye}%
  \BibitemOpen
  \bibfield  {author} {\bibinfo {author} {\bibfnamefont {D.~T.}\ \bibnamefont
  {Son}},\ }\href {\doibase 10.1103/PhysRevD.78.046003} {\bibfield  {journal}
  {\bibinfo  {journal} {Phys. Rev.}\ }\textbf {\bibinfo {volume} {D78}},\
  \bibinfo {pages} {046003} (\bibinfo {year} {2008})},\ \Eprint
  {http://arxiv.org/abs/0804.3972} {arXiv:0804.3972 [hep-th]} \BibitemShut
  {NoStop}%
\bibitem [{\citenamefont {Balasubramanian}\ and\ \citenamefont
  {McGreevy}(2008)}]{Balasubramanian:2008dm}%
  \BibitemOpen
  \bibfield  {author} {\bibinfo {author} {\bibfnamefont {K.}~\bibnamefont
  {Balasubramanian}}\ and\ \bibinfo {author} {\bibfnamefont {J.}~\bibnamefont
  {McGreevy}},\ }\href {\doibase 10.1103/PhysRevLett.101.061601} {\bibfield
  {journal} {\bibinfo  {journal} {Phys. Rev. Lett.}\ }\textbf {\bibinfo
  {volume} {101}},\ \bibinfo {pages} {061601} (\bibinfo {year} {2008})},\
  \Eprint {http://arxiv.org/abs/0804.4053} {arXiv:0804.4053 [hep-th]}
  \BibitemShut {NoStop}%
\bibitem [{\citenamefont {Herzog}\ \emph {et~al.}(2008)\citenamefont {Herzog},
  \citenamefont {Rangamani},\ and\ \citenamefont {Ross}}]{Herzog:2008wg}%
  \BibitemOpen
  \bibfield  {author} {\bibinfo {author} {\bibfnamefont {C.~P.}\ \bibnamefont
  {Herzog}}, \bibinfo {author} {\bibfnamefont {M.}~\bibnamefont {Rangamani}}, \
  and\ \bibinfo {author} {\bibfnamefont {S.~F.}\ \bibnamefont {Ross}},\ }\href
  {\doibase 10.1088/1126-6708/2008/11/080} {\bibfield  {journal} {\bibinfo
  {journal} {JHEP}\ }\textbf {\bibinfo {volume} {0811}},\ \bibinfo {pages}
  {080} (\bibinfo {year} {2008})},\ \Eprint {http://arxiv.org/abs/0807.1099}
  {arXiv:0807.1099 [hep-th]} \BibitemShut {NoStop}%
\bibitem [{\citenamefont {Maldacena}\ \emph {et~al.}(2008)\citenamefont
  {Maldacena}, \citenamefont {Martelli},\ and\ \citenamefont
  {Tachikawa}}]{Maldacena:2008wh}%
  \BibitemOpen
  \bibfield  {author} {\bibinfo {author} {\bibfnamefont {J.}~\bibnamefont
  {Maldacena}}, \bibinfo {author} {\bibfnamefont {D.}~\bibnamefont {Martelli}},
  \ and\ \bibinfo {author} {\bibfnamefont {Y.}~\bibnamefont {Tachikawa}},\
  }\href {\doibase 10.1088/1126-6708/2008/10/072} {\bibfield  {journal}
  {\bibinfo  {journal} {JHEP}\ }\textbf {\bibinfo {volume} {0810}},\ \bibinfo
  {pages} {072} (\bibinfo {year} {2008})},\ \Eprint
  {http://arxiv.org/abs/0807.1100} {arXiv:0807.1100 [hep-th]} \BibitemShut
  {NoStop}%
\bibitem [{\citenamefont {Adams}\ \emph {et~al.}(2008)\citenamefont {Adams},
  \citenamefont {Balasubramanian},\ and\ \citenamefont
  {McGreevy}}]{Adams:2008wt}%
  \BibitemOpen
  \bibfield  {author} {\bibinfo {author} {\bibfnamefont {A.}~\bibnamefont
  {Adams}}, \bibinfo {author} {\bibfnamefont {K.}~\bibnamefont
  {Balasubramanian}}, \ and\ \bibinfo {author} {\bibfnamefont {J.}~\bibnamefont
  {McGreevy}},\ }\href {\doibase 10.1088/1126-6708/2008/11/059} {\bibfield
  {journal} {\bibinfo  {journal} {JHEP}\ }\textbf {\bibinfo {volume} {0811}},\
  \bibinfo {pages} {059} (\bibinfo {year} {2008})},\ \Eprint
  {http://arxiv.org/abs/0807.1111} {arXiv:0807.1111 [hep-th]} \BibitemShut
  {NoStop}%
\bibitem [{\citenamefont {Hartnoll}\ and\ \citenamefont
  {Herzog}(2008)}]{Hartnoll:2008hs}%
  \BibitemOpen
  \bibfield  {author} {\bibinfo {author} {\bibfnamefont {S.~A.}\ \bibnamefont
  {Hartnoll}}\ and\ \bibinfo {author} {\bibfnamefont {C.~P.}\ \bibnamefont
  {Herzog}},\ }\href {\doibase 10.1103/PhysRevD.77.106009} {\bibfield
  {journal} {\bibinfo  {journal} {Phys. Rev.}\ }\textbf {\bibinfo {volume}
  {D77}},\ \bibinfo {pages} {106009} (\bibinfo {year} {2008})},\ \Eprint
  {http://arxiv.org/abs/0801.1693} {arXiv:0801.1693 [hep-th]} \BibitemShut
  {NoStop}%
\bibitem [{\citenamefont {Adams}\ and\ \citenamefont
  {Yaida}(2011)}]{Adams:2011rj}%
  \BibitemOpen
  \bibfield  {author} {\bibinfo {author} {\bibfnamefont {A.}~\bibnamefont
  {Adams}}\ and\ \bibinfo {author} {\bibfnamefont {S.}~\bibnamefont {Yaida}},\
  }\href@noop {} {\  (\bibinfo {year} {2011})},\ \Eprint
  {http://arxiv.org/abs/1102.2892} {arXiv:1102.2892 [hep-th]} \BibitemShut
  {NoStop}%
\bibitem [{\citenamefont {Peet}\ and\ \citenamefont
  {Polchinski}(1999)}]{Peet:1998wn}%
  \BibitemOpen
  \bibfield  {author} {\bibinfo {author} {\bibfnamefont {A.~W.}\ \bibnamefont
  {Peet}}\ and\ \bibinfo {author} {\bibfnamefont {J.}~\bibnamefont
  {Polchinski}},\ }\href {\doibase 10.1103/PhysRevD.59.065011} {\bibfield
  {journal} {\bibinfo  {journal} {Phys. Rev.}\ }\textbf {\bibinfo {volume}
  {D59}},\ \bibinfo {pages} {065011} (\bibinfo {year} {1999})},\ \Eprint
  {http://arxiv.org/abs/hep-th/9809022} {arXiv:hep-th/9809022 [hep-th]}
  \BibitemShut {NoStop}%
\end{thebibliography}

%

\end{document}